\documentclass[10pt,letterpaper,twocolumn]{article}

\usepackage[letterpaper,
  top=0.98in,
  bottom=0.98in,
  left=0.98in,
  right=0.98in,
  headheight=22pt,
  headsep=8pt,
  footskip=24pt]{geometry}

\usepackage[T1]{fontenc}
\usepackage{newtxtext,newtxmath} % Times-like text and mathematics
\usepackage{microtype}
\usepackage{graphicx}
\usepackage{booktabs}
\usepackage{array}
\usepackage{amsmath}
\usepackage{enumitem}
\usepackage{titlesec}
\usepackage[table]{xcolor}
\usepackage{natbib}
\usepackage{fancyhdr}
\usepackage{lastpage}
\usepackage{caption}
\usepackage{url}
\usepackage{hyperref} 
\hypersetup{colorlinks,citecolor=blue,linkcolor=blue,urlcolor=blue}
\usepackage[dvipsnames]{xcolor}
 
\definecolor{customColor}{RGB}{230, 230, 230}
\definecolor{customColor1}{RGB}{204, 255, 204}
\definecolor{customColor2}{RGB}{255, 217, 178}

\newcommand{\IACPaperCode}{IAC-26,A4,1,5,x109607}
\newcommand{\IACTitle}{Opportunity Lost in Searching for Technosignatures}
\newcommand{\IACConference}{77\textsuperscript{th} International Astronautical Congress (IAC 2026), Antalya, T\"urkiye, 5--9 Oct 2026.}
\newcommand{\IACCopyright}{Copyright \textcopyright 2026 by the International Astronautical Federation (IAF). All rights reserved.}

\titleformat{\section}
  {\normalfont\bfseries\fontsize{10}{11}\selectfont}
  {\thesection.}{0.45em}{}
\titlespacing*{\section}{0pt}{9pt}{1pt}

\titleformat{\subsection}
  {\normalfont\itshape\fontsize{10}{11}\selectfont}
  {\thesubsection}{0.45em}{}
\titlespacing*{\subsection}{0pt}{7pt}{1pt}

\titleformat{\subsubsection}
  {\normalfont\itshape\fontsize{10}{11}\selectfont}
  {\thesubsubsection}{0.45em}{}
\titlespacing*{\subsubsection}{0pt}{6pt}{1pt}

\newcommand{\iacunnumbered}[1]{%
  \par\vspace{7pt}%
  \noindent\textbf{#1}\par\vspace{1pt}}

\setlist{nosep,leftmargin=*}

\begin{document}

% ============================================================
% Full-width title, author information, abstract, and keywords
% ============================================================
\twocolumn[{%
\begin{minipage}{\textwidth}
\begin{center}
  \fontsize{11}{12}\selectfont
  \IACPaperCode

  \vspace{8pt}

  {\bfseries\fontsize{12}{14}\selectfont \IACTitle}

  \vspace{8pt}

  {\bfseries\fontsize{11}{13}\selectfont
  Steve Prabu\textsuperscript{1,2,*}, David DeBoer\textsuperscript{1,3}, Andrew Siemion\textsuperscript{1,3,4}, Andre R. Offringa\textsuperscript{5}, Natasha Hurley-Walker\textsuperscript{2}, Charlie K. Ashe\textsuperscript{6}, Sofia. Z. Sheikh\textsuperscript{3,4}, Steve Croft\textsuperscript{1,3,4}, Greg Sleap\textsuperscript{2}}
\end{center}

\vspace{5pt}

\noindent\textsuperscript{1}\textit{University of Oxford, Department of Physics, Astrophysics, Denys Wilkinson Building, Keble Road, OX1 3RH Oxford, United Kingdom}\par
\vspace{-1pt}
\noindent\textsuperscript{2}\textit{International Centre for Radio Astronomy Research -- Curtin University, Perth, WA 6845, Australia}\par
\vspace{-1pt}
\noindent\textsuperscript{3}\textit{Radio Astronomy Lab, University of California, 501 Campbell Hall, Berkeley 94720, USA}\par
\vspace{-1pt}
\noindent\textsuperscript{4}\textit{SETI Institute, 339 Bernardo Ave, Suite 200, Mountain View, CA 94043, USA}\par
\vspace{-1pt}
\noindent\textsuperscript{5}\textit{Netherlands Institute for Radio Astronomy (ASTRON), PO Box 2, NL-7990 AA Dwingeloo, the Netherlands}\par
\vspace{-1pt}
\noindent\textsuperscript{6}\textit{School of Physics, Trinity College Dublin, College Green, Dublin 2, D02 PN40, Ireland}\par
\vspace{-1pt}

\vspace{8pt}

\noindent * Corresponding Author: Steve Prabu: steve.prabu@physics.ox.ac.uk

\vspace{11pt}

\begin{center}
\textbf{Abstract}
\end{center}
\vspace{-4pt}

\noindent
The sensitivity of radio technosignature searches has improved dramatically over the last five decades; however, as the RFI environment from which these searches are conducted has become increasingly polluted, our advancements in telescope sensitivity have not necessarily been complemented by less ambiguous technosignature searches. In this work, we investigate the resulting loss of opportunity to conduct unambiguous technosignature searches from Earth. We introduce three new concepts, \textsf{opportunity lost}, \textsf{volume lost}, and \textsf{self-censorship}, to quantify the impact of orbiting RFI sources on Earth-based technosignature searches. Using publicly available orbital data between 2008 and 2025, we find an approximately two-fold increase in the spatial occupancy of satellites within representative telescope beams, with the exact magnitude of the loss strongly dependent on telescope pointing. We further demonstrate that the cumulative effect of faint emission from multiple satellites increase the effective system temperature and consequently reduce the volume over which a technosignature of a given intrinsic power can be detected. Using approximately 10,000 archival observations from the Murchison Widefield Array spanning 2014--2024, we find suggestive evidence for increasing RFI occupancy across most of the frequencies observed by the MWA. While other opportunities for terrestrial technosignature searches exist, we argue for the protection and utilisation of uniquely radio-quiet environments such as the lunar farside, which offers an opportunity to regain the ability to perform unambiguous technosignature searches that is being progressively lost on Earth.

\vspace{2pt}

\vspace{8pt}
\end{minipage}
}]

% ============================================================
% Main two-column manuscript
% ============================================================

% \iacunnumbered{Nomenclature}
% This section is not numbered. A nomenclature section could be provided when there are mathematical symbols in your paper. Superscripts and subscripts must be listed separately. Nomenclature definitions should not appear again in the text.

% \iacunnumbered{Acronyms/Abbreviations}
% This section is not numbered. Define acronyms and abbreviations that are not standard in this section. Such acronyms and abbreviations that are unavoidable in the abstract must be defined at their first mention there. Ensure consistency of abbreviations throughout the article. Always use the full title followed by the acronym (abbreviation) to be used, e.g., reusable suborbital launch vehicle (RSLV), International Space Station (ISS).

\section{Introduction}

One of our most promising opportunities to unambiguously answer the question ``Are we alone?'' is the use of radio telescopes to search for technosignatures \citep{Tarter2007}. The detection of a credible technosignature originating beyond Earth would not only provide compelling evidence for the existence of extraterrestrial life, but would also strongly suggest that the discovered life form is intelligent and capable of developing technologies that can produce the detected technosignature. One of the prominent unexplained signals encountered in the history of technosignature searches is the ``Wow!'' signal \citep{wow}, detected by the Big Ear radio telescope in 1977. Despite being a technosignature candidate, this signal remains in part unexplained due to the limited observational information available at that time, making it challenging to determine the signal's origin. Since then, despite enormous advances in observational and computational capabilities over the last five decades, no extraterrestrial technosignature has been identified. Over this period, collecting areas of radio telescopes have increased by over an order of magnitude\footnote{$3,000~\mathrm{m^2}$ for the 64-m Parkes radio telescope built in the 1960s \citep{CSIROParkes}, compared to a collecting area of $32,000~\mathrm{m^2}$ for the full SKA-Mid \citep{SKAOSciencePerformance}.}, accompanied by substantial advances in detection algorithms and computational power\footnote{From approximately $10^8$ FLOP s$^{-1}$ for the Cray-1 supercomputer in the 1970s to $10^{13}$--$10^{14}$ FLOP s$^{-1}$ for a single modern GPU.}. However, alongside these advances, the anthropogenic radio-frequency environment has become increasingly complex. Thus, while our ability to perform sensitive technosignature searches and characterise candidate signals has improved dramatically, the growing prevalence of technosignature-like RFI introduces an increasing source of ambiguity. Given the profound implications of claiming a technosignature detection, and it helping us answer the question ``Are we alone?'' and understanding our plance in the Universe, such a detection carries an extraordinary evidentiary burden and a need to be unambiguous. This motivates the central premise of this work: to investigate how the evolving terrestrial RFI environment is progressively reducing the parameter space within which technosignature searches can be conducted unambiguously.\\

Paradoxically, the same technological progress that has increased our ability to search for technosignatures has also transformed the electromagnetic environment from which those searches are conducted. Technologies developed to make human life safer, more convenient, entertaining, and interconnected have surrounded terrestrial radio observatories with an increasingly complex background of anthropogenic radio-frequency interference (RFI) \citep{Offringa2015, Hobbs2020}. In effect, as humanity has become more electromagnetically conspicuous to a hypothetical distant observer, we have simultaneously made it more difficult for ourselves to listen unambiguously for faint technological signals from elsewhere in the Universe. Satellite-based transmitters are particularly problematic in this regard. Unlike many terrestrial sources of RFI, satellite RFI signals originate from the sky and can exhibit several characteristics associated with candidate technosignatures, including narrow bandwidths, time-dependent Doppler drift, polarisation, modulation, and an apparent origin on the sky \citep[e.g.,][]{sheikh2021analysis}. Furthermore, while individual satellite systems may occupy relatively restricted orbital configurations, the rapid proliferation of large mega-constellations distributed across multiple orbital planes and inclinations substantially expands the range of sky positions and Doppler signatures over which anthropogenic signals may appear in radio astronomy datasets. Consequently, distinguishing a human-made signal from a genuine extraterrestrial technosignature can become increasingly difficult. While RFI mitigation and subtraction techniques may recover otherwise compromised astronomical data, the presence of technosignature-like RFI can nevertheless introduce residual ambiguity that is particularly consequential given the evidentiary burden associated with a technosignature detection. Hence, the increasing prevalence of satellite-based RFI presents a growing challenge to conducting unambiguous Earth-based technosignature searches.\\

In this paper, we develop new metrics and a framework to explore this increasing opportunity lost. In Section~\ref{sec:framework}, we define three new metrics and a pipeline framework that we use to quantify the opportunity lost. We discuss the implications of our results for present and future technosignature searches in Section~\ref{sec:results}, before summarising our conclusions in Section~\ref{sec:conclusion}.\\

\section{Framework, Definitions, and Results}
\label{sec:framework}
To investigate the diminishing opportunity for unambiguous technosignature searches from Earth, we introduce three concepts: \textsf{opportunity lost}, \textsf{volume lost}, and \textsf{self-censorship}. We further introduce an empirical framework, based on flagging statistics from radio-astronomical datasets, to characterise the increasing prevalence of RFI in the terrestrial observing environment. These concepts are defined below.\\

\subsection{Opportunity Lost ($L_{O}$) and Volume Lost ($L_{V}$)}

In the current era of satellite megaconstellations, approximately 4,500 objects were launched into space in the previous year\footnote{\url{https://ourworldindata.org/grapher/yearly-number-of-objects-launched-into-outer-space}}, with further accelerated\footnote{with space based data-centres being proposed for the near future \url{https://www.theguardian.com/science/2026/jul/23/space-datacenters-bezos-blue-origin}} growth expected in the coming years. Consequently, the prevalence of satellite-generated signals with characteristics resembling candidate technosignatures in radio-astronomical datasets is increasing. This growing population reduces the opportunity to conduct technosignature searches that are unambiguously distinguishable from anthropogenic signals. Here, we define two metrics, \textsf{opportunity lost} ($L_{O}$) and \textsf{volume lost} ($L_{V}$), to quantify two distinct ways in which human-made satellites can reduce the effectiveness of terrestrial technosignature searches.\\

The first metric, \textsf{opportunity lost} ($L_{O}$), captures the fraction of synthesised beam pointings within the telescope's primary beam that are occupied by satellites at a given epoch and hence are unusable for unambiguous technosignature searches\footnote{In the following, we use the term primary beam to refer to the beam response of the individual antenna in the interferometer, and the term synthesised beam to refer to the interferometric beam pattern generated by $n$ antenna elements together.}. The value of $L_{O}$ therefore depends on the satellite population at the epoch of observation, the sizes of the telescope's primary beam and synthesised beam, the geographical location of the observatory, and the telescope pointing direction. The metric $L_{O}$ classifies a synthesised beam pointing within the telescope's primary beam as unusable for an unambiguous technosignature search if it spatially overlaps with a satellite's position. As the emission properties of all satellites are unknown, and the emission properties of individual satellites can evolve with time (as the electronics onboard deviate from their initial design), we adopt the conservative assumption that such a pointing is no longer usable for an unambiguous technosignature search, given the weight of the question ``Are we alone?'' addressed by technosignature searches. In contrast, \textsf{volume lost} ($L_{V}$) quantifies the reduction in effective search volume resulting from an elevated system noise floor produced by the cumulative contribution of many weak satellite emissions. An increase in the effective noise floor reduces the sensitivity of a technosignature search and, consequently, the volume of space over which a transmitter of a given intrinsic power could be detected. The derivation of $L_{V}$ is presented in Appendix A, and the final equation is given below. \\

\begin{equation}
    L_{V} = 1 - \left( \frac{T_{0}}{T_{0} + \Delta T} \right)^{3/2},
    \label{Eq:main:1}
\end{equation}

where $T_{0}$ is the noise temperature of the telescope in an RFI-free environment, and $\Delta T$ is the increase in temperature contributed by the cumulative emission from all satellites within the telescope's beam. If there are $n$ satellites within the beam, with each satellite at a line of sight distance $d_i$ from the telescope, then the increase in temperature $\Delta T$ is given by

\begin{equation}
    \Delta T = \frac{A_{\rm eff}}{k_{B}} \times \sum_{i=1}^{n} \frac{{\rm EIRP}_{i\nu}}{4\pi d_{i}^2},
    \label{Eq:main:2}
\end{equation}

where $A_{\rm eff}$ is the effective collecting area of the telescope, $k_{B}$ is the Boltzmann constant, and ${\rm EIRP}_{i\nu}$ is the EIRP per unit bandwidth of the $i$th satellite. We direct the reader to Appendix A for more information. \\

To examine the evolution of these metrics, we obtained the entire public Two-Line Element (TLE) archive published by the 18th Space Defense Squadron and made available through Space-Track\footnote{\url{https://www.space-track.org/}}. Although the complete available archive was downloaded (with the earliest entries dating to 2004), we restrict our analysis to TLEs from 2008 onwards. Earlier epochs contain substantial inconsistencies in the number of satellites for which TLEs were made publicly available. During the years 2004--2008, TLEs for many satellites that had previously been published were no longer made publicly available in subsequent epochs due to the satellites being deemed classified. \\

\begin{figure*}
    \centering
    \includegraphics[width=1\linewidth]{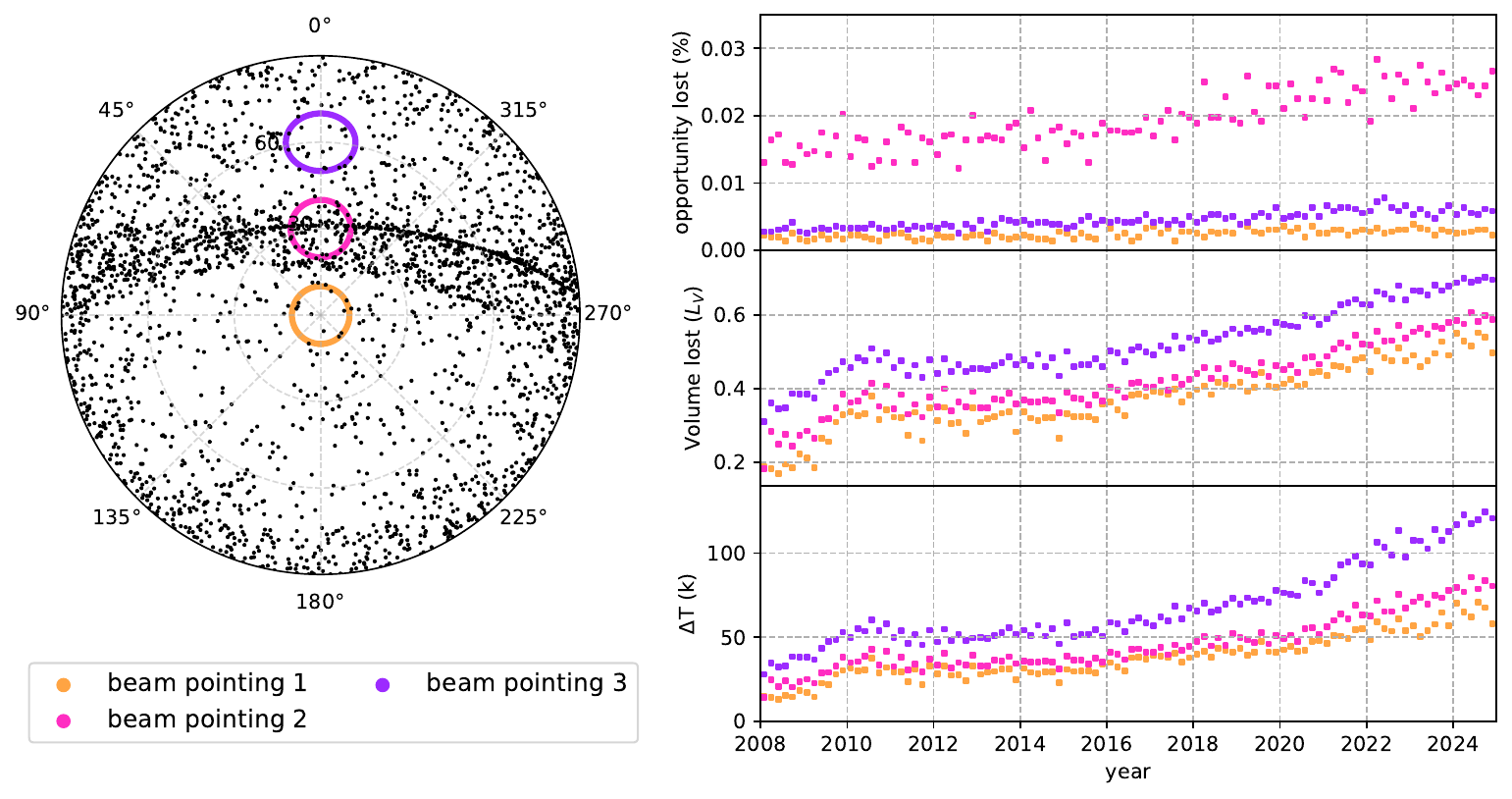}
    \caption{Figure shows the progressive loss of opportunity to perform unambiguous technosignature searches from Earth. The left panel shows all unclassified satellites visible from a chosen location on Earth, computed for an arbitrarily selected day in 2025, along with three different telescope beam pointings. For each of these beams, the corresponding \textsf{opportunity lost}, \textsf{volume lost}, and increase in system temperature are shown in the right panels. }
    \label{fig:oppLost}
\end{figure*}

Using the publicly available TLEs from 2008 onwards, we propagate the orbits of satellites from 2008--2025. In Figure~\ref{fig:oppLost}, we show the evolution of the \textsf{opportunity lost} and \textsf{volume lost} metrics for a representative observatory location and telescope beam size, chosen to correspond to those of the Murchison Widefield Array \citep[MWA;][]{tingay2013murchison, wayth2018phase, tingay2026mwa3}, for which observational data are considered later in this work. In the left panel of Figure~\ref{fig:oppLost}, we show all visible unclassified satellites (as seen by the MWA) for an arbitrarily chosen epoch in 2025. We also show three different beam pointings with a $10^{\circ}$ full-width at half-maximum, with one beam pointing at zenith, one beam pointing $30^{\circ}$ due north, and one beam pointing $60^{\circ}$ due north along the meridian. These beams are referred to as Beam~1, Beam~2, and Beam~3, respectively, throughout the rest of the paper. \\

For the three beams, using Equations \ref{Eq:main:1} and \ref{Eq:main:2}, we calculate the \textsf{volume lost} fraction and the corresponding increase in system temperature over time, as shown in the middle and bottom-right panels of Figure~\ref{fig:oppLost}. To model the \textsf{volume lost} and increase in system temperature, we conservatively assume that all satellites within the telescope beam have an EIRP of $1$~pW~Hz$^{-1}$ (either due to downlink transmissions, unintended emissions, or reflection of terrestrial transmisions like FM). When integrated over a MHz bandwidth, this remains comparable to some known satellite reflections of terrestrial FM transmissions \citep{Prabu2020}, and significantly lower than measured Starlink UEMR at VHF frequencies \citep{grigg2023starlink}. The evolution and implications of these metrics are discussed further in Section~\ref{sec:discussionOpplLost}.

\subsection{Empirical Evidence for Opportunity Lost}

Radio telescopes are often constructed in remote locations to minimise the impact of anthropogenic RFI. One such instrument is the MWA, a low-frequency, wide-field radio interferometer co-located with SKA-Low at Inyarrimanha Ilgari Bundara, the CSIRO Murchison Radio-astronomy Observatory in remote Western Australia, one of the most radio-quiet observatory sites on Earth. Despite its location within a protected radio-quiet environment, the MWA has not escaped interference from orbiting transmitters \citep{Prabu2020, Prabu2020b, prabu2022sensitivity, prabu2023orbit, hennessy2020orbit, hennessy2022establishing}. Here, using flagging statistics obtained from approximately $10,000$ archival MWA observations spanning 2014--2024, we investigate the temporal evolution of RFI occupancy in radio-astronomical datasets. \\

The MWA operates between $80-300$\,MHz, with an instantaneous bandwidth of $30.72$\,MHz and an electronically steerable, wide-field primary beam. The population of RFI sources detected by a radio telescope is strongly dependent on pointing direction. Observations towards lower elevations can be more susceptible to terrestrial RFI, whose prevalence may additionally vary with azimuth, whereas observations towards zenith preferentially sample the overhead population of orbiting RFI sources. To minimise these directional dependencies and enable a consistent comparison across epochs, we restrict our analysis to only night-time\footnote{Small variations in the Sun during the day can result in very different flagging and it could prove challenging to isolate the flagging due to RFI and flagging due to Sun in the day observations. } MWA zenith drift-scan observations.\\

We identified approximately 58,000 unique zenith drift-scan observations from the MWA archive\footnote{MWA All-Sky Virtual Observatory (ASVO): \url{https://asvo.mwatelescope.org/}} for which more than 100 antenna elements (`tiles') were operational. From this sample, we selected $9594$ observations\footnote{The observation IDs and the pipelines used can be obtained from the supplementary section.}, corresponding to approximately 330 hours of data, to obtain approximately uniform coverage across MWA observing frequency and epoch over the ten-year period. Perfectly uniform sampling was not possible because these data were repurposed from archival observations, for which observing strategies, frequency coverage, cadence, and scientific objectives evolved over time. The resulting temporal and frequency coverage is shown in APPENDIX Figure~\ref{fig:mwacampaign}. The lower panels of APPENDIX Figure~\ref{fig:mwacampaign} additionally show the different array configurations and corresponding baseline distributions represented within the selected observations. More information on the MWA observations is provided in APPENDIX B. \\

\begin{figure*}
    \centering
    \includegraphics[width=1\linewidth]{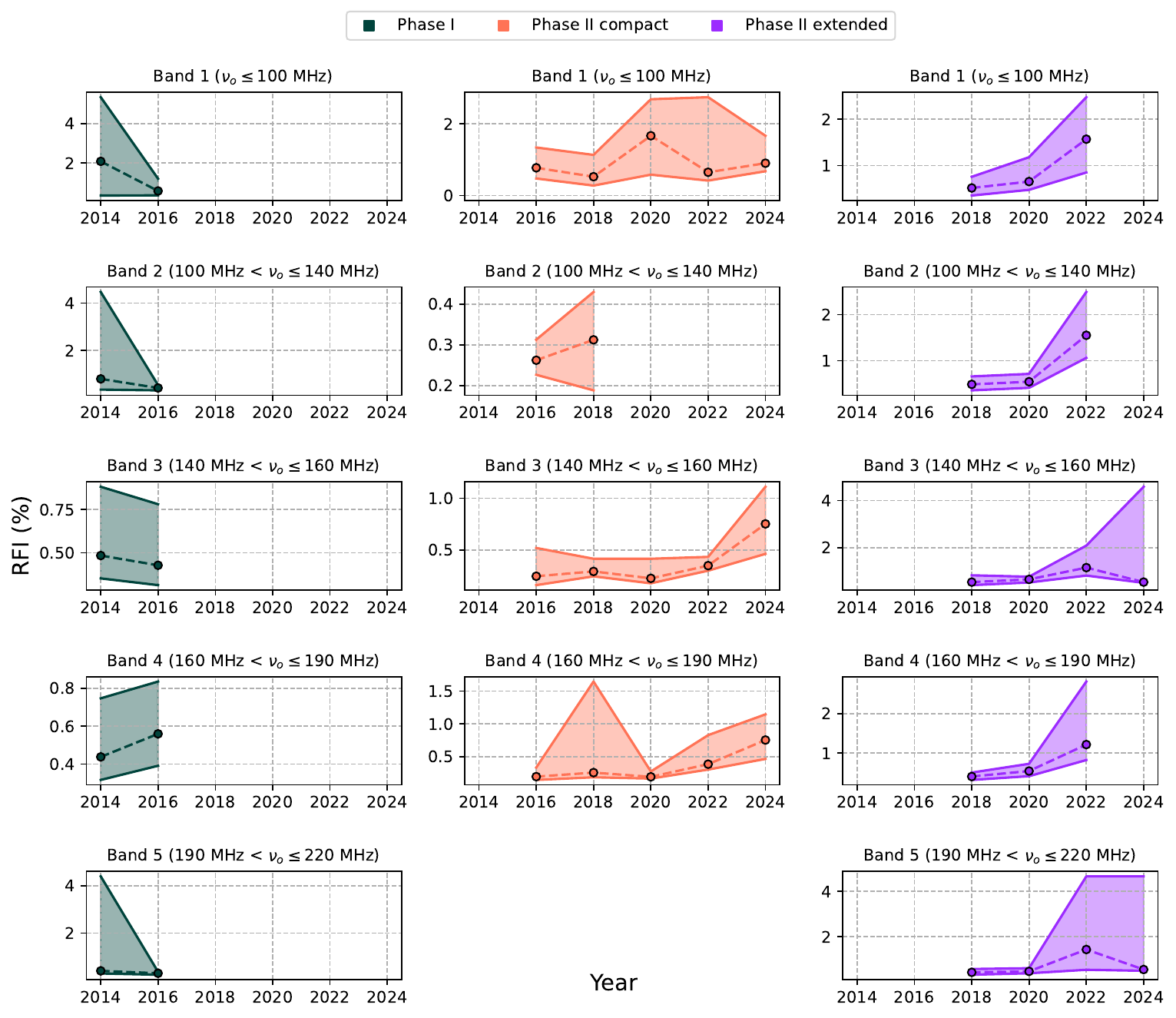}
    \caption{for the different MWA configurations (Phase I, Phase II compact, and Phase II extended), we show the temporal evolution of the percentage of data flagged by {\tt AOFLAGGER}, using the default MWA flagging strategy, across the five different frequency bands over the 10 years for which archival data were obtained. The circle marker shows the median flagging percentage, while the shaded region shows the 25th--75th percentile range of the data, binned every two years. }
    \label{fig:mwaflagging}
\end{figure*}

\begin{figure*}
    \centering
    \includegraphics[width=1\linewidth]{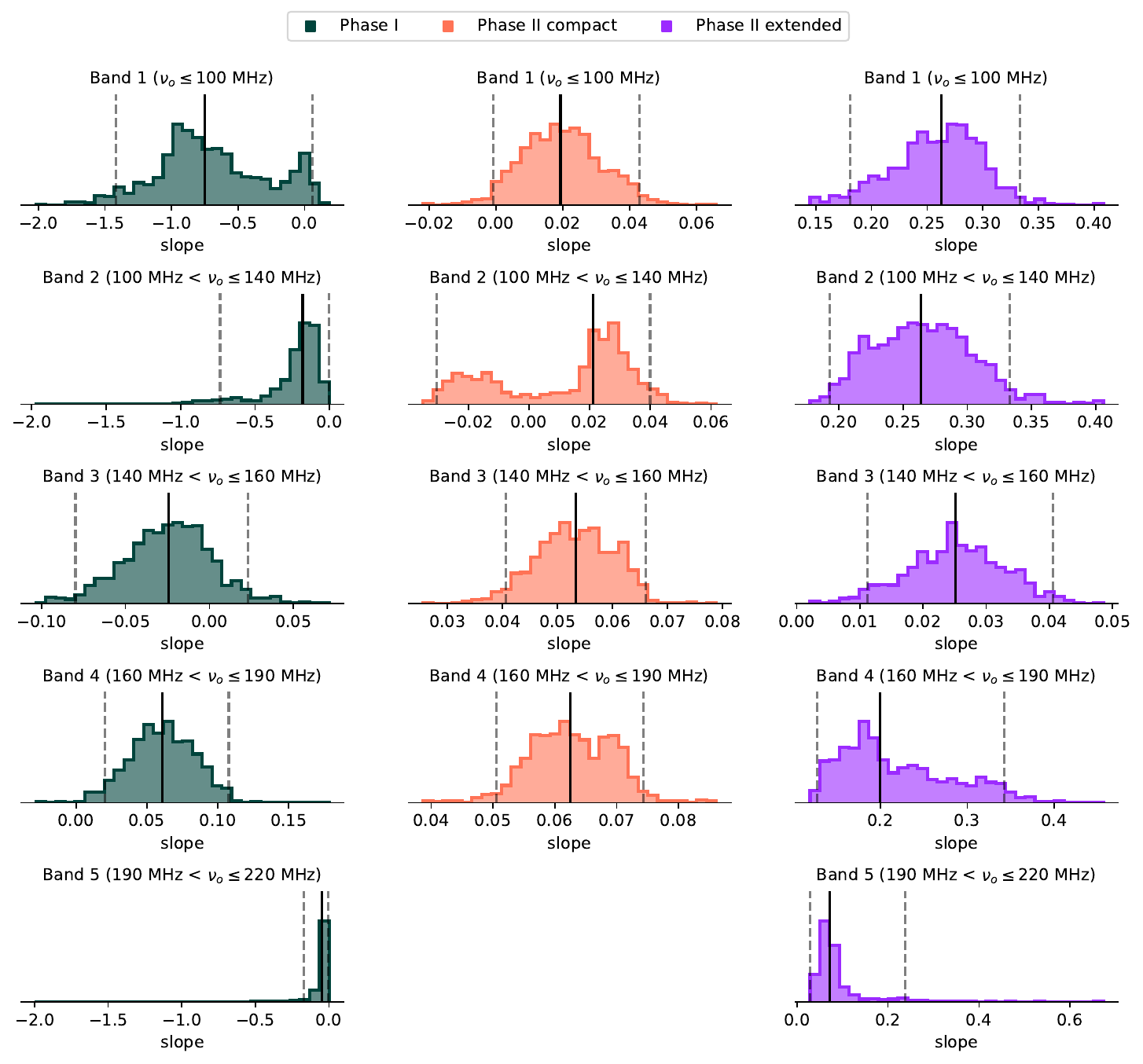}
    \caption{Distribution of slopes from the bootstrap regression fit for each of our band+array combination. In the above panels, a positive slope indicates increasing RFI occupancy with time, a zero slope suggests no discernible trend within the available precision of the data, and a negative slope suggests decreasing RFI occupancy with time. In vertical solid line we show the median of the distribution, and using vertical dashed lines we show the $99\%$ high density interval.  }
    \label{fig:mwaflaggingKS}
\end{figure*}

Several previous studies have demonstrated the prevalence of satellite-based RFI in MWA and other radio-astronomical datasets \citep[e.g.][]{Prabu2020b, grigg2025enhanced, grigg2025impact, grigg2023starlink}. These studies generally employ analysis pipelines specifically optimised for detecting satellite signals, including techniques such as shift-stacking \citep{prabu2022sensitivity, grigg2022dug} and the use of known satellite ephemeris to increase the signal-to-noise ratio of weak satellite emissions. While highly effective for identifying individual satellites, such techniques can be computationally expensive when applied to large archival datasets. For the present analysis, we instead use the widely adopted radio-frequency-interference flagging package {\tt AOFLAGGER}\footnote{\url{https://aoflagger.readthedocs.io/en/latest/}} to process the selected MWA observations and record their flagging statistics, as a computationally cheaper proxy for capturing the impact of satellite based RFI\footnote{We note that {\tt AOFLAGGER} can also capture terrestrial RFI and not just orbiting RFI. However, as we do not expect significant increase in terrestrial RFI over time. The temporal evolution in RFI occupancy is expected to be dominated by satellite-based RFI.}. As {\tt AOFLAGGER} is routinely applied as the first step in most MWA science pipelines (except in some transient studies), the resulting flagged fraction provides a practical empirical measure of the fraction of otherwise usable astronomical data rendered unavailable by RFI. Its evolution with time therefore provides a complementary empirical measure of the increasing RFI occupancy (or opportunity lost for unambiguous technosignature searches) considered within the framework developed in this work.\\

\begin{table*}[h!]
\caption{Summary of the linear regression slopes obtained by fitting a linear model to the temporal RFI trends for different MWA array configurations and frequency bands. In green \fcolorbox{black}{customColor1}{\rule{0pt}{1pt}\rule{1pt}{0pt}}, we show band/configuration combinations with a significant increasing trend (i.e., zero lies outside the $99\%$ highest density interval [HDI]); in orange \fcolorbox{black}{customColor2}{\rule{0pt}{1pt}\rule{1pt}{0pt}}, we show significant decreasing trends; and in grey \fcolorbox{black}{customColor}{\rule{0pt}{1pt}\rule{1pt}{0pt}}, we show band/configuration combinations with no discernible increasing or decreasing trend. }
\centering

\renewcommand{\arraystretch}{1.5}
\begin{tabular*}{\textwidth}{@{\extracolsep{\fill}} |c|c|c|c|c|c| @{}}
\hline
Array Configuration & Band 1 & Band 2 & Band 3 & Band 4 & Band 5 \\
\hline

Phase I 
& \cellcolor{gray!20}$-0.76_{-0.91}^{+0.84}$
& \cellcolor{gray!20}$-0.18_{-0.78}^{+0.19}$
& \cellcolor{gray!20}$-0.03_{-0.07}^{+0.06}$
& \cellcolor{gray!20}$0.06_{-0.07}^{+0.06}$
& \cellcolor{gray!20}$-0.04_{-0.4}^{+0.05}$ \\
\hline

Phase II Compact 
& \cellcolor{gray!20}$0.02_{-0.03}^{+0.03}$
& \cellcolor{gray!20}$0.02_{-0.05}^{+0.04}$
& \cellcolor{green!20}$0.05_{-0.02}^{+0.01}$
& \cellcolor{green!20}$0.06_{-0.02}^{+0.02}$
& NA \\
\hline

Phase II Extended 
& \cellcolor{green!20}$0.27_{-0.12}^{+0.1}$
& \cellcolor{green!20}$0.26_{-0.08}^{+0.12}$
& \cellcolor{green!20}$0.03_{-0.02}^{+0.02}$
& \cellcolor{green!20}$0.20_{-0.08}^{+0.19}$
& \cellcolor{green!20}$0.07_{-0.04}^{+0.45}$ \\
\hline

\end{tabular*}

\label{tab:linearFitSummary}
\end{table*}

For each MWA observation, we obtain flagging statistics using the default MWA flagging strategy implemented within the {\tt AOFLAGGER} software. We explore the use of a more aggressive flagging strategy in APPENDIX B. A similar application of {\tt AOFLAGGER} to characterise the RFI environment of the MWA was presented by \citet{Offringa2015}, 10 years ago. Here, we extend that analysis by examining the temporal evolution of RFI occupancy over a substantially longer time baseline, using a larger number of observations and higher frequency resolution.\\

The dominant sources and characteristics of RFI vary substantially within MWA frequencies; for example, the FM broadcast band can contain satellite-reflections of  terrestrial FM, while the region around $136$\,MHz is strongly affected by ORBCOMM satellite downlinks. To avoid combining physically distinct RFI populations, we divide the processed observations into five frequency bands and analyse their flagging statistics separately. Observations with central frequencies $\leq100$\,MHz are assigned to Band~1, $100$--$140$\,MHz to Band~2, $140$--$160$\,MHz to Band~3, $160$--$190$\,MHz to Band~4, and $190$--$220$\,MHz to Band~5. Relatively few suitable MWA observations were available above $220$\,MHz, owing partly to reduced instrumental sensitivity and increased calibration challenges associated with primary-beam aliasing \citep{Cook2021}, and we therefore exclude these higher frequencies from the present analysis. The frequency boundaries were selected to broadly separate the dominant expected RFI populations. Band~1 predominantly samples satellite reflections of terrestrial FM transmissions; Band~2 encompasses prominent satellite downlinks, including ORBCOMM; Band~3 includes frequencies potentially affected by reflected digital television (DTV) transmissions from satellites and aircraft; while Bands~4 and~5 sample the remaining spectrum considered in this analysis.\\

The temporal evolution of the percentage of MWA data being flagged by {\tt AOFLAGGER} is shown in Figure~\ref{fig:mwaflagging}. We note that three different MWA configurations were present within the chosen observations (MWA Phase~I array, MWA Phase~II compact array, and MWA Phase~II extended array), and as these three arrays have different ground footprints and potentially sample the RFI environment differently, we show the flagging statistics for each configuration separately. As a first-order test to determine whether the temporal evolution of RFI for a given band/array configuration is increasing, decreasing, or stationary, we perform a linear regression for each array and frequency combination. We note that we do not assert that a linear model describes the temporal behaviour of RFI, but instead use the slope of the linear fit as a simple measure of whether the RFI occupancy is increasing, decreasing, or stationary (or whether a trend is indiscernible within the precision of the available data).\\

To perform the regression, we adopt a bootstrap approach in which a linear model is fitted to the median flagging percentage as a function of time. We note that the regression requires an estimate of the uncertainty in the measured median, rather than the 25th--75th percentile ranges shown in Figure~\ref{fig:mwaflagging}. The latter captures the spread in the range of RFI occupancy present within the dataset, rather than the uncertainty in the median flagging percentage. For each time bin within an array+band combination, we randomly draw 100 samples, calculate their median, and perform a linear fit to the resulting temporal trend. We repeat this random sampling and fitting procedure 1,000 times, producing a distribution of possible slopes for each array + band combination, as shown in Figure~\ref{fig:mwaflaggingKS}. We also tabulate the median and $99\%$ highest density interval (HDI) of each slope distribution in Table~\ref{tab:linearFitSummary}. Of the 14 combinations of band and array configuration investigated, seven show positive slopes, suggesting increasing trends, and seven show slopes consistent with zero, indicating no discernible trend within the precision of the available data. We note that none of the band+array combination provided a negative slope that did not overlap with 0 within the $99\%$ HDI. As all the array+band combinations show either a positive slope (increasing RFI occupancy) or a slope consistenat with zero, a likely interpretation is an overall increasing occupancy of RFI with time in radio-astronomy datasets, resulting in increasing fractnal data being lost to flagging, and consequently opportunities for technosignature searches being lost. \\

\subsection{Opportunity Lost due to Self-Censorship}

Astronomers often deliberately avoid observing in frequency ranges known to contain persistent or strong RFI as a mitigation strategy and to preserve data quality. While scientifically justified, this practice introduces what we term \textsf{self-censorship}: the a priori exclusion of portions of the accessible electromagnetic spectrum due to expected anthropogenic interference. For technosignature searches, this represents an opportunity lost to discover signals occupying these frequencies. \\

\begin{figure*}
    \centering
    \includegraphics[width=0.9\linewidth]{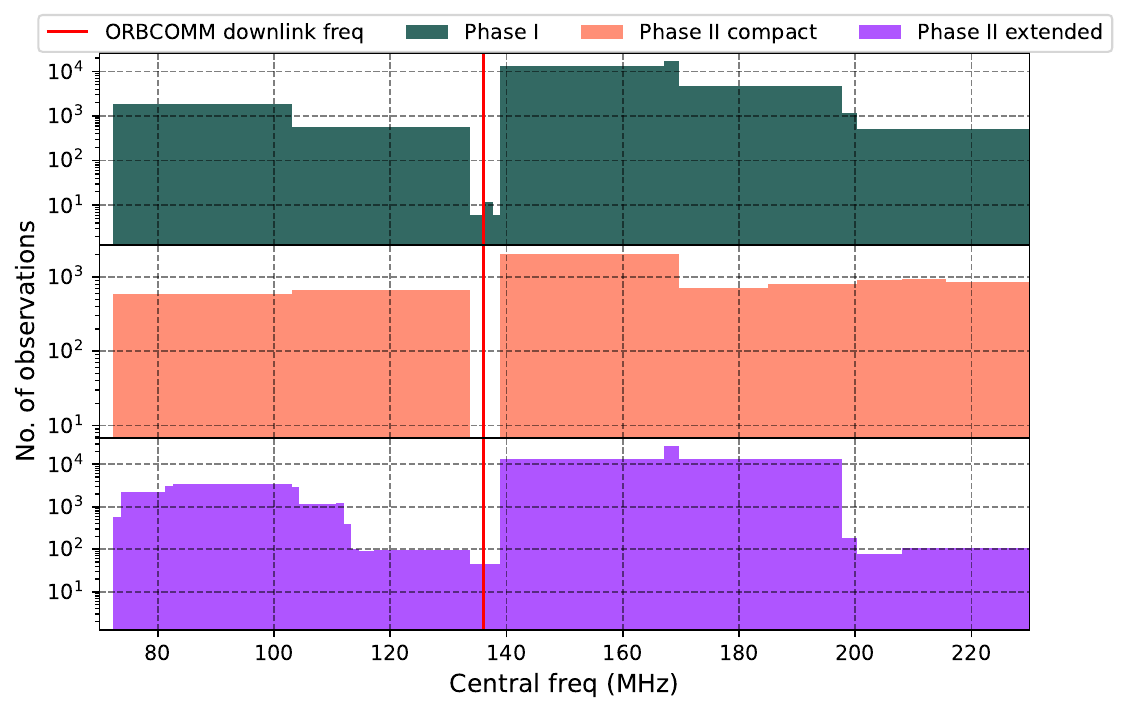}
    \caption{Figure illustrates the opportunity lost due to \textsf{self-censorship}. The figure shows the frequency coverage of approximately 50,000 observations identified from the MWA archive as night-time zenith drift scans with more than 100 operational tiles. We observe a clear deficit in the number of observations at the ORBCOMM downlink frequencies, demonstrating the opportunity lost to search for technosignatures at these frequencies.}
    \label{fig:censorship}
\end{figure*}

One example of \textsf{self-censorship} at MWA frequencies is the ORBCOMM satellite downlink band. Figure \ref{fig:censorship} shows the observing frequencies of the $\sim58,000$ zenith drift-scan observations identified from the MWA archive\footnote{This represents the complete sample of suitable zenith drift-scan observations with more than 100 operational tiles, rather than the subset used for the RFI flagging analysis.}, revealing a clear deficit of observations around the ORBCOMM frequencies. The few observations in \ref{fig:censorship} that observe in ORBCOMM frequencies are few experiments that specifically targeted ORBCOMM, such as for mapping out the MWA beam using OBRCOMM passes \citep{Chokshi2021}. This effect is particularly relevant to technosignature experiments employing commensal backends \citep{CzechMeerKAT, TremblayVLA, PriceParkes}, where frequency choices made for the primary science programme directly determine the available technosignature search space. ORBCOMM provides one example of such \textsf{self-censorship} at VHF frequencies, with similar losses potentially occurring at higher-frequency satellite-navigation and communication bands. \

\section{Discussions}
\label{sec:results}

\subsection{Opportunity Lost On Earth}
\label{sec:discussionOpplLost}

Through the introduction of the \textsf{opportunity lost}, \textsf{volume lost}, and \textsf{self-censorship} metrics, we demonstrate the diminishing opportunity to conduct unambiguous technosignature searches from Earth. Our analysis of \textsf{opportunity lost} shows an approximately two-fold increase in the spatial occupancy of satellites within the primary beam of an MWA-like telescope between 2008 and 2025 (Figure \ref{fig:oppLost}), progressively reducing the fraction of telescope pointings free from ambiguity (in the context of technosignature searches). Techniques such as near-field RFI subtractions \citep{Prabu_Tingay_Williams_2023, finlay2023trajectory} can be used to model and remove satellite signals from interferometric data. While these approaches can mitigate satellite RFI for many astronomical science cases, technosignature searches present a fundamentally different challenge. For example, an astrophysical transient (for example, a Fast Radio Burst [FRB])  detected following the subtraction of satellite RFI can be independently identified through characteristic astrophysical properties (such as its dispersion for an FRB, which also is not impossible for a satellite to mimic through Doppler changes as it drifts through the telescope's beam). In contrast, satellite RFI can exhibit many of the same characteristics sought in candidate technosignatures, including narrow bandwidth, Doppler drift, modulation, and an apparent origin on the sky \citep[e.g,][]{sheikh2021analysis}. Consequently, a technosignature candidate recovered from data containing a modelled and subtracted satellite signal would carry substantially greater ambiguity. Moreover, because the search for technosignatures directly addresses the question “Are we alone?", and thereby helps us understand our own place in the Universe, there is an extraordinary evidentiary burden associated with claiming the detection of a technosignature and a corresponding need for such a detection to be unambiguous. Hence, RFI mitigation/subtraction techniques may recover otherwise compromised astronomical data, but cannot necessarily recover the opportunity to perform an \textit{unambiguous} technosignature search.\\

Our \textsf{volume lost} analysis demonstrates that the cumulative contribution of low-level RFI emissions from many satellites can have a significant impact on the effective noise floor of a radio telescope and, consequently, on the volume lost relative to an RFI-free observing environment. An elevated noise floor reduces the maximum distance over which a technosignature of a given EIRP can be detected, resulting in a loss of opportunity to probe more distant technosignatures that would otherwise have been detectable. Using the satellite emission model described in the previous section and a collecting area of $5200$\,$m^2$ (comparable to modern data radio interferometer), from Figure \ref{fig:oppLost}, we note that the cumulative contribution of even tens of satellites can raise the effective noise floor by approximately $10^{\circ}-20^{\circ}$K. We note that our analysis does not account for satellites outside the primary beam, which can number in the hundreds and may contribute additional power through the attenuated sidelobe response of the instrument. As the received power is directly proportional to the collecting area of the telescope, we expect SKA-like instruments to experience greater \textsf{volume lost} fraction due to their significantly larger collecting areas. Parallel studies with the Engineering Development Array have shown the occupancy of Satellite signals in current day radio astronomy datasets \citep{2026arXiv260901995T} \\

 From the right panels of Figure~\ref{fig:oppLost}, we find that the increase in system noise, \textsf{opportunity lost}, and \textsf{volume lost} vary for different beam pointings. The zenith-pointed beam (Beam~1) has the lowest \textsf{opportunity lost} due to the lower density of satellites within the beam. Lower-elevation beams (Beam~2 and Beam~3) contain a larger number of satellites within the primary beam, as they probe a larger projected volume of the sky. Beam~2 sees the highest density of satellites within the primary beam, as it intersects the geostationary belt, which contains an overdensity of satellites, as seen in the left panel of Figure~\ref{fig:oppLost}. Hence, to minimise \textsf{opportunity lost} for technosignature searches, we recommend zenith or near-zenith pointings, as they are statistically more likely to contain fewer satellites within the beam, and hence lesser ambiguity from humanmade RFI sources. We note that this may not hold true for observatories located near the equator, where the geostationary belt passes close to the zenith and may therefore result in greater \textsf{opportunity lost}. However, a zenith-pointed technosignature search is not always possible, as not all pointings carry equal weight for technosignature searches, particularly for targeted technosignature searches of known candidate star systems. A similar trend is seen for \textsf{volume lost}, where lower-elevation pointings receive greater cumulative power from satellites, resulting in a larger increase in antenna temperature and, consequently, greater \textsf{volume lost}. We note that, as we only use unclassified objects whose TLEs are publicly available, our analysis provides a lower limit on the true \textsf{opportunity lost} and \textsf{volume lost}, as it does not capture contributions from classified objects.\\

From Figure \ref{fig:mwaflagging} and \ref{fig:mwaflaggingKS}, we find  evidence for increasing occupancy of RFI in radio telescope datasets, resulting in a loss of opportunity for technosignature searches.  In Band~1, where we suspect an increased number of specular reflections of terrestrial FM transmissions, we find a marginal increase in the flagging percentage in 2024 compared to 2018 and the pre-Starlink era. However, in Band~2 (which contains ORBCOMM and Starlink downlink frequencies \citep{grigg2025impact}),  Band~4 and Band~5, we find strong evidence for an increased flagging percentage, with upto and order of magnitude increase in the percentage of data being flagged in some bands over the last decade.\\

\subsection{Opportunity found?}
Despite the continuous opportunity to conduct unambiguous technosignature searches being lost on Earth, other environments, such as the lunar farside, provide opportunities to search for technosignatures without ambiguity arising from RFI. The lunar farside is the only location within the inner Solar System that is currently RFI-free and within cost-effective access. Because the Moon is tidally locked to the Earth, the lunar farside is permanently shielded from Earth-based RFI and RFI sources in the near-Earth environment. The effectiveness of this shielding was demonstrated by the Radio Astronomy Explorer 2 (RAE-2) \citep{alexander1975scientific}, which measured a substantial reduction in terrestrial radio emission while passing behind the Moon. However, the increasing accessibility of cislunar space may mean that this uniquely radio-quiet environment cannot be assumed to remain so indefinitely. Recent simulations suggest that unintended electromagnetic radiation (UEMR) from future lunar orbiters, landers, and surface infrastructure could progressively affect the radio quietness of the lunar farside over the next few years \citep{charlieLunar}. Hence, missions such as the Lunar Farside Transients and Technology Telescope \citep[LFT3;][]{daveLFT3} aim to make the most of this unique opportunity provided by the pristine environment of the lunar farside to perform unambiguous technosignature searches before it becomes contaminated by UEMR from lunar orbiters and landers. In the context of orbiting RFI, the current lunar farside is found to be comparable to Earth in the 1960s (with a similar number of orbiting RFI sources), an environment much cleaner than when the ``Wow!'' signal or pulsars were discovered \citep{daveLFT3}. \\

The terrestrial experience considered in this work provides a potential lesson for the future use of the lunar farside. Scientific instruments, communications infrastructure, navigation systems, and future human activities on and around the Moon will inevitably place competing demands on the lunar electromagnetic environment. Many of these activities provide substantial scientific and societal benefits, but without appropriate spectrum management and electromagnetic-compatibility considerations, they may also progressively reduce the radio-quiet characteristics that make the lunar farside uniquely valuable. The lunar farside may therefore represent not simply an opportunity found, but a finite opportunity that requires deliberate protection. Preserving portions of this environment for radio astronomy and technosignature searches could prevent the terrestrial pattern described in this work (where increasing technological capability is accompanied by a diminishing ability to conduct unambiguous searches) from being repeated on the Moon.\\

\subsection{Other Opportunity Persists on Earth} 
Despite the increasing ambiguity associated with technosignature searches from Earth due to the growing occupancy of satellite-based RFI within radio observations, opportunities to conduct technosignature searches from terrestrial facilities still persist. Technosignatures can broadly arise in two ways. First, intelligent life elsewhere in the Universe may transmit powerful deliberate beacons, analogous to our own Arecibo planetary radar transmissions \citep{Nolan2026}. Such signals may be comparatively easy to detect but transient in nature. Due to the increasing prevalence of transient technosignature-like RFI, our opportunity to unambiguously identify such signals from Earth is becoming progressively reduced. However, intelligent civilisations may also produce electromagnetic signatures as an unintended consequence of their everyday technological activities, analogous to terrestrial FM transmitters, airport radars, satellite navigation systems, and other communication infrastructure. Although these unintended technosignatures are likely to be substantially fainter than deliberate beacons, they may also be more persistent. Their persistence provides repeated opportunities for detection, follow-up, and verification, potentially allowing them to be distinguished from transient terrestrial or satellite-based RFI. Additional observing strategies, such as the use of widely spaced coherent detectors (DeBoer et al., in prep.), which employ spatially separated receiving elements that do not share the same population of orbiting RFI sources, may further reduce ambiguity in technosignature searches. Thus, while the parameter space available for unambiguous technosignature searches from Earth is increasingly constrained, it is far from exhausted, and continued development of observing and verification strategies can preserve valuable opportunities for terrestrial searches.\\

\section{Conclusion}
\label{sec:conclusion}
In this work, we have introduced the concepts of \textsf{opportunity lost}, \textsf{volume lost}, and \textsf{self-censorship} to quantify different ways in which the growing anthropogenic RFI environment is reducing the parameter space available for unambiguous technosignature searches from Earth. Using the historical evolution of the satellite population, we demonstrate the progressive loss of opportunity, firstly due to the increased occupancy of satellites within a telescope's beam, and secondly due to the increased contribution of satellite emissions to the telescope noise floor. From approximately 10,000 archival observations performed by the Murchison Widefield Array, we find increasing trends in the fraction of data being flagged from 2018 onwards across all frequency bands between $70$--$220$\,MHz, coinciding with the commencement of large satellite megaconstellation deployments. Through the archival MWA observations, we also demonstrate how \textsf{self-censorship}, whereby astronomers purposefully avoid certain bands known to contain RFI, impacts technosignature searches through the loss of opportunities to search for technosignatures at these frequencies. Our modelling of \textsf{opportunity lost} and \textsf{volume lost} shows that zenith-pointed telescope observations (unless the observatory is located near the equator) are least impacted by satellite signals, with increasing impact for off-zenith pointings due to the larger projected volume, and hence increased number of satellites, viewed by the telescope at lower elevations. Nevertheless, despite the opportunity being progressively lost for unambiguous searches for deliberate technosignature beacons from elsewhere in the Universe, opportunities remain to search for persistent technosignatures arising from unintended emissions analogous to our own terrestrial FM transmissions. At the same time, the comparatively pristine lunar farside provides a rare opportunity to conduct searches with substantially reduced anthropogenic ambiguity. The terrestrial experience therefore provides an important lesson: improvements in telescope sensitivity alone do not necessarily translate into improvements in our ability to search unambiguously, and preserving radio-quiet environments may be as important as developing increasingly sensitive instruments in our effort to answer the question, ``Are we alone?''

\iacunnumbered{Acknowledgements}
This scientific work uses data obtained from Inyarrimanha Ilgari Bundara / the Murchison Radio-astronomy Observatory. We acknowledge the Wajarri Yamaji People as the Traditional Owners and native title holders of the Observatory site. Establishment of CSIRO's Murchison Radio-astronomy Observatory is an initiative of the Australian Government, with support from the Government of Western Australia and the Science and Industry Endowment Fund. Support for the operation of the MWA is provided by the Australian Government (NCRIS), under a contract to Curtin University administered by Astronomy Australia Limited. This work was supported by resources provided by the Pawsey Supercomputing Research Centre with funding from the Australian Government and the Government of Western Australia.  SP acknowledges Breakthrough Listen (funded by the Breakthrough Prize foundation) for their Breakthrough Listen fellowship. SP would also like to thank Steven Tingay for the valuable discussions. \\

\iacunnumbered{Data Availability}
The list of MWA observations used in this study can be obtained from our Zenodo repository \url{https://zenodo.org/records/22725053}, and the MWA processing pipeline can be obtained from our Github repository \url{https://github.com/StevePrabu/MWA-RFIwatch}.\\

\iacunnumbered{Appendix A. Volume lost}
\label{sec:volumeLost}
If a satellite at a distance $d$ has an EIRP of $L_{\rm EIRP}$, then a radio telescope with an effective collecting area $A_{\rm eff}$ receives a total power $P$, given by

\begin{equation}
    P = L_{\rm EIRP} \times \frac{1}{4\pi d^2} \times A_{\rm eff}.
\end{equation}

For an observing bandwidth $B$, the power received by the radio telescope can then be converted to a change in temperature, $\Delta T$, using

\begin{equation}
    \Delta T = \frac{P}{k_B \times B},
\end{equation}

where $k_B$ is the Boltzmann constant. Combining the above two equations, we obtain the following expression relating the increase in telescope temperature to the satellite EIRP per unit bandwidth, ${\rm EIRP}_{\nu}$, which is assumed to be $1$~pW~Hz$^{-1}$ in this work:

\begin{equation}
    \Delta T = \frac{{\rm EIRP}_{\nu}}{k_B} \times \frac{A_{\rm eff}}{4\pi d^2}.
    \label{eq1}
\end{equation}

The minimum detectable flux density, $S_{\rm min}$, of a radio telescope with noise temperature $T_{\circ}$ and integration time $t$ is given by

\begin{equation}
    S_{\rm min} \propto \frac{T_{\circ}}{A_{\rm eff}\sqrt{Bt}}.
\end{equation}

The flux density from a transmitter at a distance $d$ is given by

\begin{equation}
    S_{\rm RFI} = \frac{L_{\rm EIRP}}{4\pi d^2}.
\end{equation}

Setting $S_{\rm RFI} = S_{\rm min}$, we obtain

\begin{equation}
    d_{\rm max} \propto \frac{1}{\sqrt{T_{\circ}}},
\end{equation}

where $d_{\rm max}$ is the maximum distance probed by the radio telescope for a given SNR. As the volume probed, $V$, is proportional to the third power of distance ($V \propto d_{\rm max}^3$), we obtain

\begin{equation}
    V \propto T_{\circ}^{-3/2}.
\end{equation}

If the system temperature in a pristine environment is $T_{\circ}$, and in an RFI environment the temperature rises to $T_{1}$, then the ratio of the volumes probed is given by

\begin{equation}
    \frac{V_{1}}{V_{\circ}} =
    \left(\frac{T_{\circ}}{T_{1}}\right)^{3/2}.
\end{equation}

As $T_{1} = T_{\circ} + \Delta T$ (from Equation~\ref{eq1}), the fractional volume lost ($L_{V}$) due to the increased noise floor is given by

\begin{equation}
    L_{V}
    = \frac{V_{\circ} - V_{1}}{V_{\circ}}
    = 1 - \left(\frac{T_{\circ}}{T_{1}}\right)^{3/2}
    = 1 - \left(\frac{T_{\circ}}{T_{\circ} + \Delta T}\right)^{3/2}.
\end{equation}

\iacunnumbered{Appendix B. MWA RFI statistics}

The complete list of observations used for the MWA analysis is provided in the Zenodo repository referenced in the Data Availability section. The custom {\tt AOFLAGGER} flagging script used in this work, along with the pipeline scripts, is provided in the GitHub repository. From the observations available in the MWA archive, we aimed to maximise coverage in both frequency and time when downselecting the 9,594 observations used to obtain the RFI statistics. The temporal and frequency coverage of these observations is shown in Appendix Figure~\ref{fig:mwacampaign}. We note that the selected observations span three different MWA configurations. Until 2016, the array was in the Phase~I configuration, with a maximum baseline of approximately 3~km. Since then, the array has regularly changed its physical layout between the Phase~II compact configuration, which contains most of its baselines within approximately 200~m, and the Phase~II extended configuration, with baselines extending to approximately 5~km. Different baseline lengths and orientations sample different spatial scales on the sky, and this effect is not accounted for in our analysis. For each MWA observation, we obtain flagging statistics using two different flagging strategies with the {\tt AOFLAGGER}. First, we apply the default MWA flagging strategy built within the {\tt AOFLAGGER} software, which uses the SumThreshold algorithm on individual interferometric baselines. Second, we employ a more sensitive and aggressive flagging strategy in which the visibilities from all baselines are coherently combined prior to flagging, thereby increasing sensitivity to fainter RFI signals. Comparing these two approaches allows us to investigate both the RFI that would routinely be excised during standard astronomical processing and the underlying population of weaker RFI detectable with increased sensitivity. \\

The summary of the RFI statistics collected for the different MWA bands is shown in Appendix Figures~\ref{fig:band1}, \ref{fig:band2}, \ref{fig:band3}, \ref{fig:band4}, and \ref{fig:band5}. We note several interesting trends; however, as these trends generally have reasonably large error bars, we only speculate on their possible causes and do not assert them to be significant. A much more extensive RFI statistics collection campaign would be required to confirm these trends with greater certainty. In Band~1, where the data are mostly expected to be impacted by terrestrial FM, we find that the flagging percentage decreases with baseline length. This is probably due to larger baseline separations seeing less mutual RFI, and hence the cross-correlation of long baselines effectively mitigates the impact of terrestrial FM transmitters. In Band~2, the strongest and most frequent RFI is the ORBCOMM downlink. As the ORBCOMM downlink is a sky-based point-source RFI, almost all baselines see a similar contribution from the RFI, and hence we observe an approximately uniform flagging percentage across baseline lengths in Band~2. We also note increased fraction of the data being flagged when observing near the galactic plane. \\

We note that, across all frequency bands, the baseline-summation method (the custom flagging strategy provided to {\tt AOFLAGGER}) consistently performs more aggressive flagging, as it is more sensitive to low-level RFI (such as the ones described in \cite{2023ApJ...957...78W}). The largest difference between the default flagging strategy adopted by {\tt AOFLAGGER} and the custom, more sensitive flagging strategy is seen in Band~1, where the custom flagging strategy consistently flags approximately an order of magnitude more data. In all other frequency bands, the differences between the flagging percentages of the default and custom strategies are smaller and generally overlap within the error bars. From the subplots showing the flagging percentage at fine-channel resolution, we find that all bands have less than $1\%$ of the data flagged, except at the ORBCOMM downlink frequencies, where consistently more than $80\%$ of the data are flagged, and five fine channels within the FM band, where more than $10\%$ of the data are flagged. We also show the flagging statistics as a function of beam pointings in APPENDIX Figure \ref{fig:mwaBL}, along with the location of a few bright radio sources in the southern sky. We note that flagging percentage goes up in most of the frequency bands when some of the bright sources (such as Fornax A and Centaurus A) are within the primary beam. \\

\begin{figure*}
    \centering
    \includegraphics[width=1\linewidth]{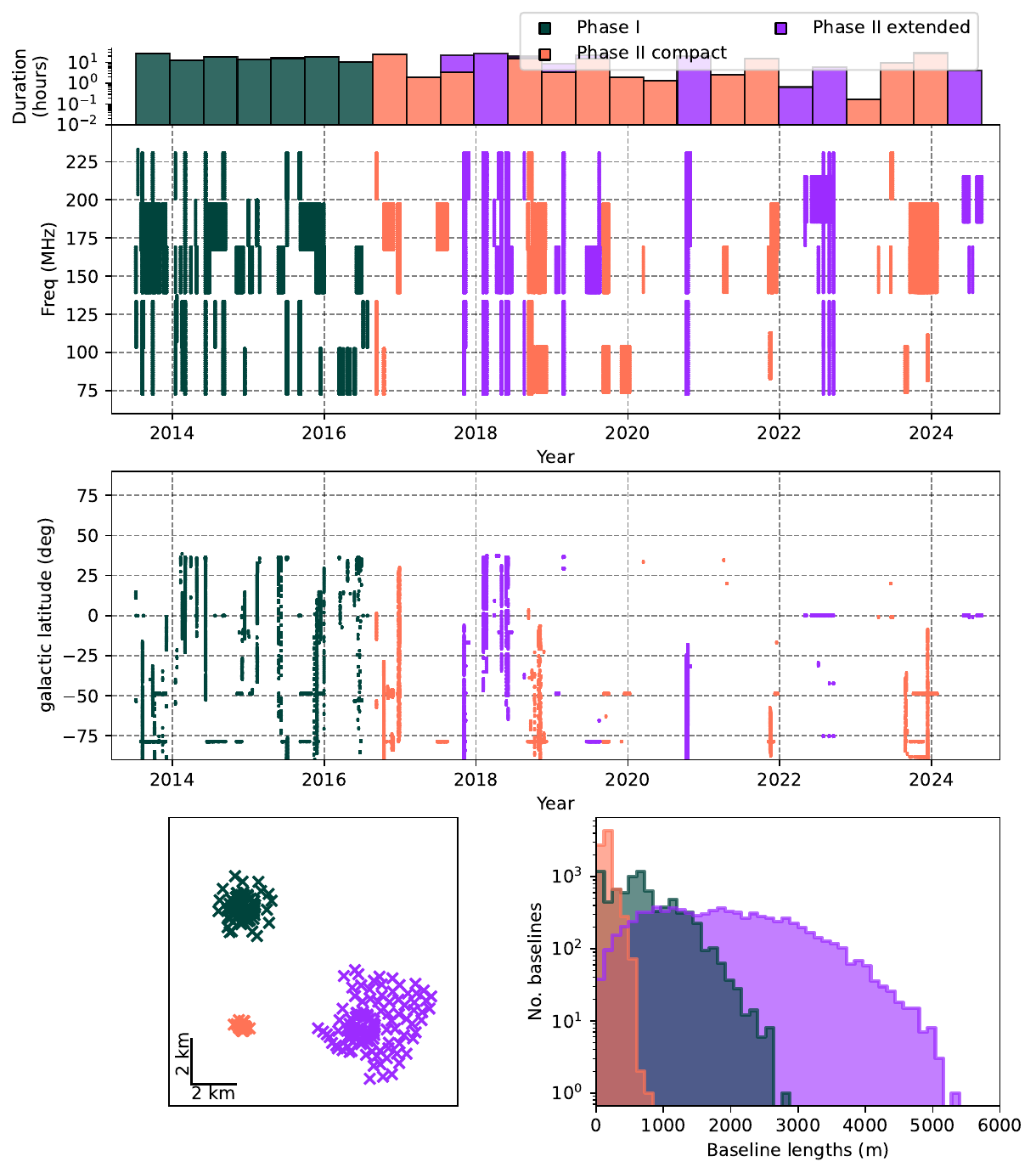}
    \caption{A high-level summary of the approximately 10,000 observations from which we collect RFI statistics in this work. The top panel shows the frequencies and epochs covered by the observations, while the middle panel shows the Galactic latitudes sampled by these observations. The bottom panels show the ground footprints of the different MWA array configurations, with the three configurations shown in different colours throughout the figure.}
    \label{fig:mwacampaign}
\end{figure*}

\begin{figure*}
    \centering
    \includegraphics[width=1\linewidth]{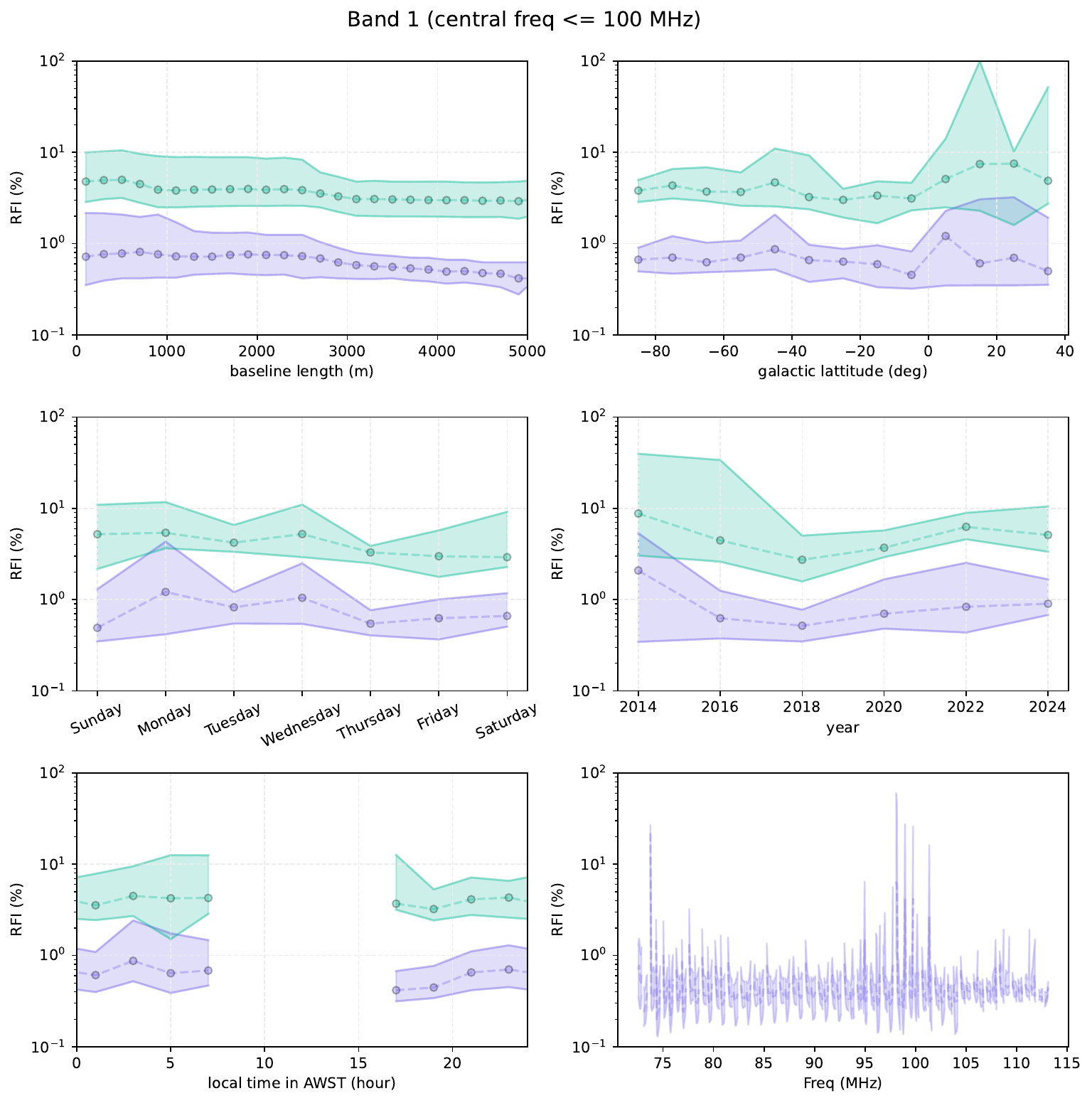}
    \caption{ Summary of the flagging statistics collected for Band~1. Purple shows the statistics obtained using the default {\tt AOFLAGGER} strategy, while teal green shows those obtained using the custom {\tt AOFLAGGER} strategy. The top-left panel shows the variation in flagging as a function of baseline length, while the top-right panel explores the variation as a function of Galactic latitude. The middle-left and bottom-left panels show the variation as a function of day of the week and hour of the day in AWST, respectively. The middle-right panel shows the variation as a function of year using all Galactic latitudes, as opposed to the restricted sample shown in Figure~\ref{fig:mwaflagging}. The bottom-right panel shows the flagging statistics at the fine-channel level. We do not show the fine-channel flagging percentage for the custom flagging strategy, as the script used to generate these statistics was heavily limited by disk I/O and the data could not be extracted within the available HPC wall time. Moreover, including both flagging strategies made the panel visually crowded; hence, we show the fine-channel flagging percentage only for the default flagging strategy.}
    \label{fig:band1}
\end{figure*}

\begin{figure*}
    \centering
    \includegraphics[width=1\linewidth]{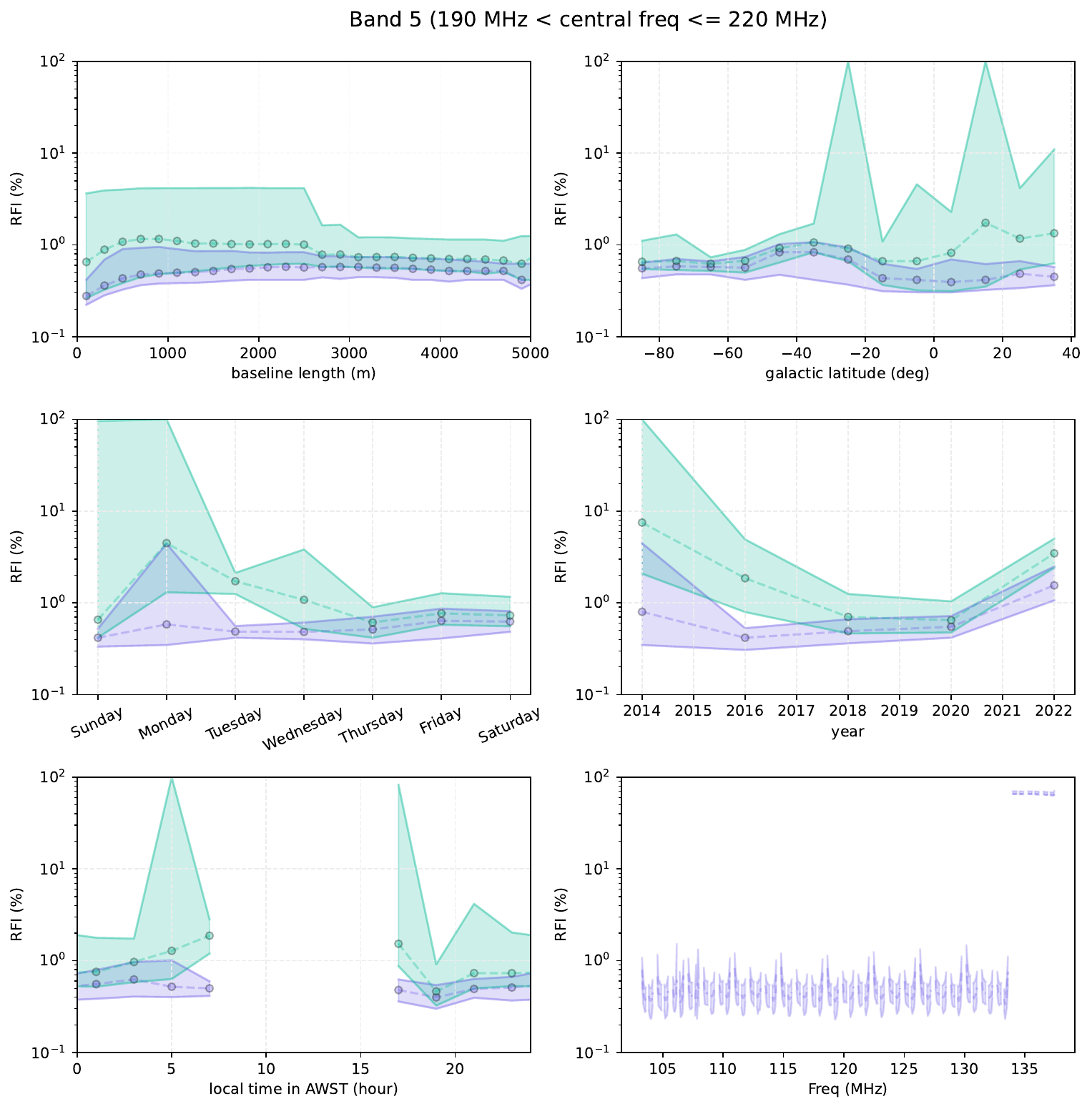}
    \caption{Same as APPENDIX Figure \ref{fig:band1}, but for Band~2}
    \label{fig:band2}
\end{figure*}

\begin{figure*}
    \centering
    \includegraphics[width=1\linewidth]{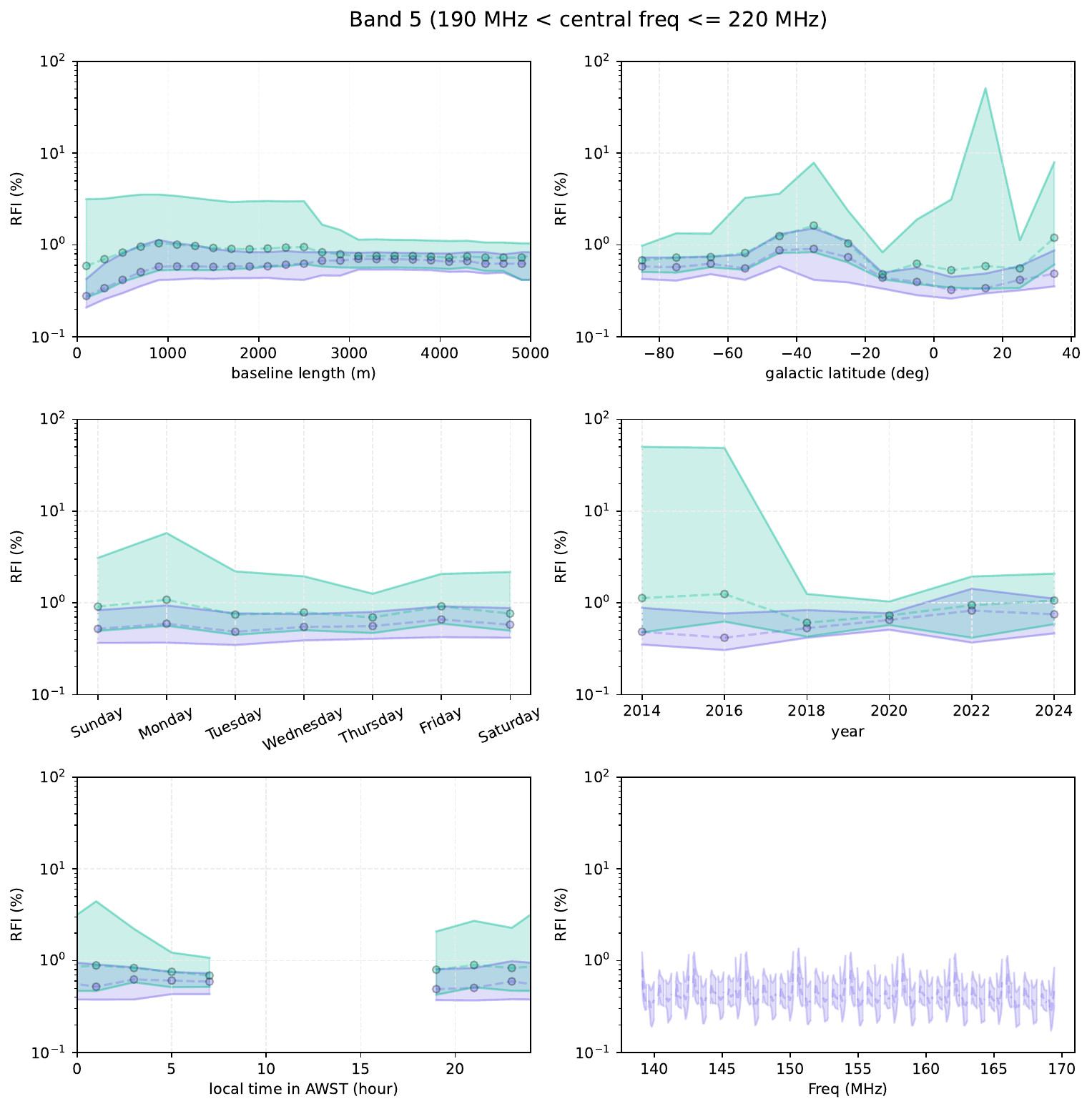}
    \caption{Same as APPENDIX Figure \ref{fig:band1}, but for Band~3}
    \label{fig:band3}
\end{figure*}

\begin{figure*}
    \centering
    \includegraphics[width=1\linewidth]{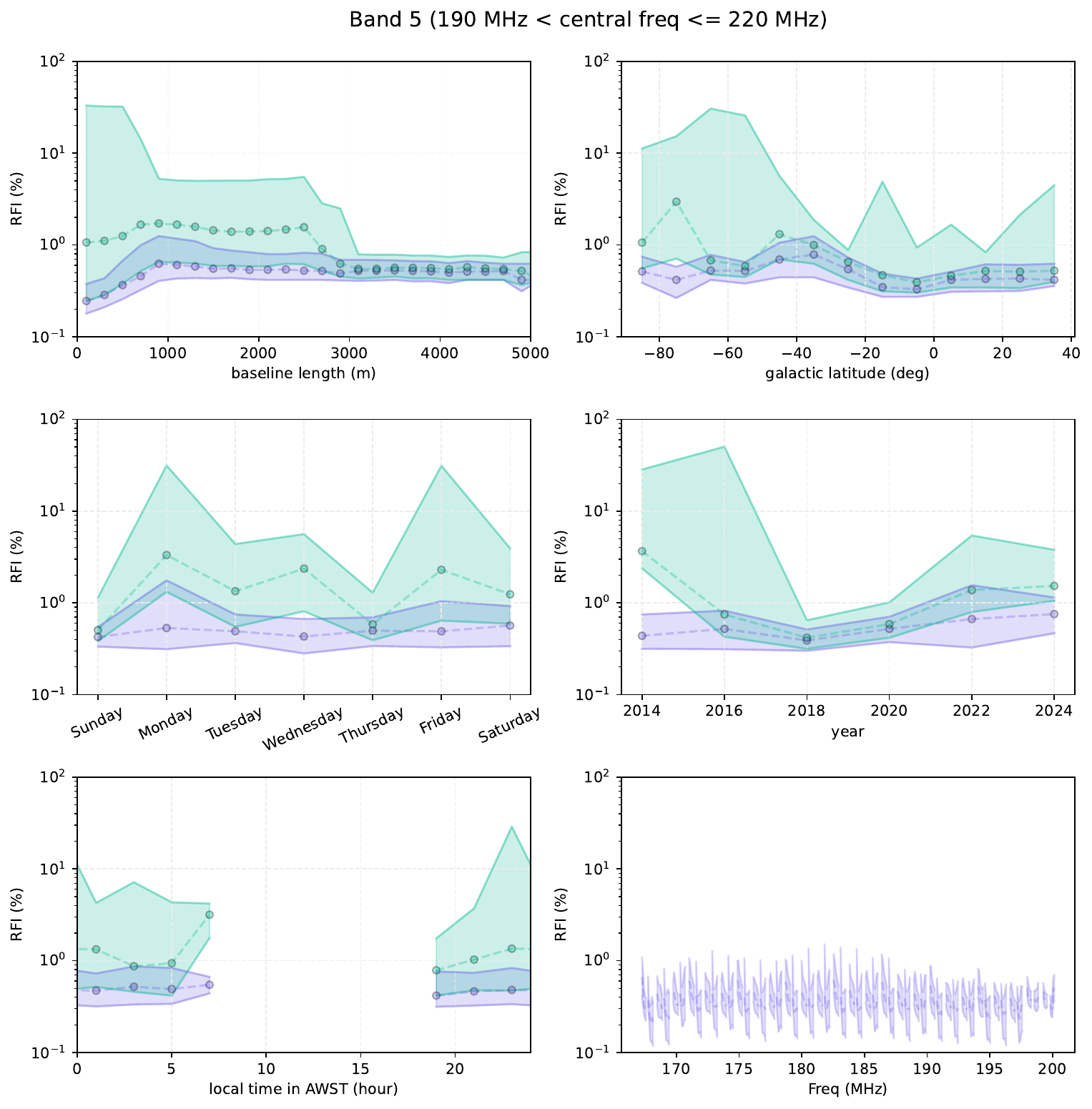}
    \caption{Same as APPENDIX Figure \ref{fig:band1}, but for Band~4}
    \label{fig:band4}
\end{figure*}

\begin{figure*}
    \centering
    \includegraphics[width=1\linewidth]{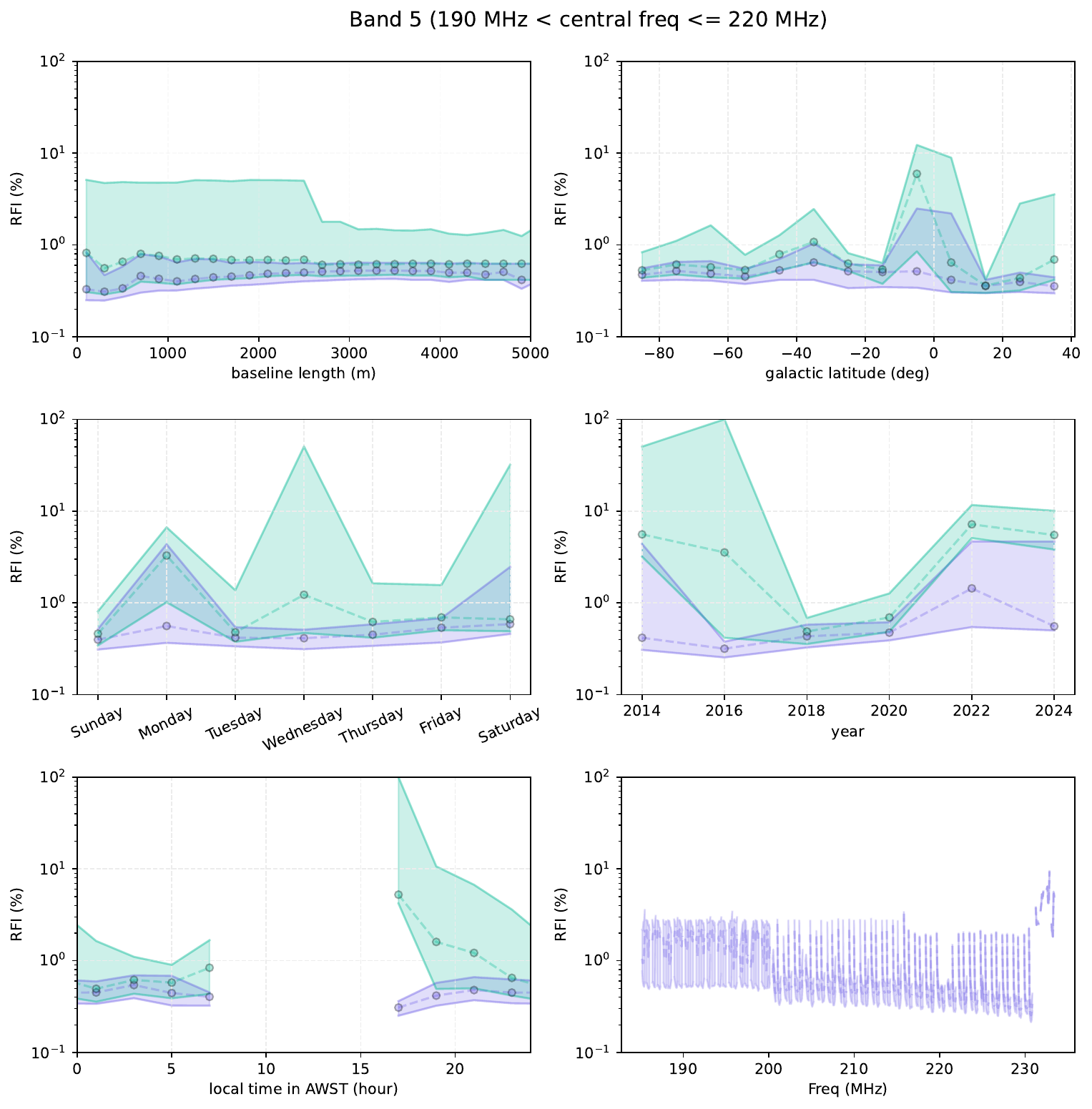}
    \caption{Same as APPENDIX Figure \ref{fig:band1}, but for Band~5}
    \label{fig:band5}
\end{figure*}

\begin{figure*}
    \centering
    \includegraphics[width=1\linewidth]{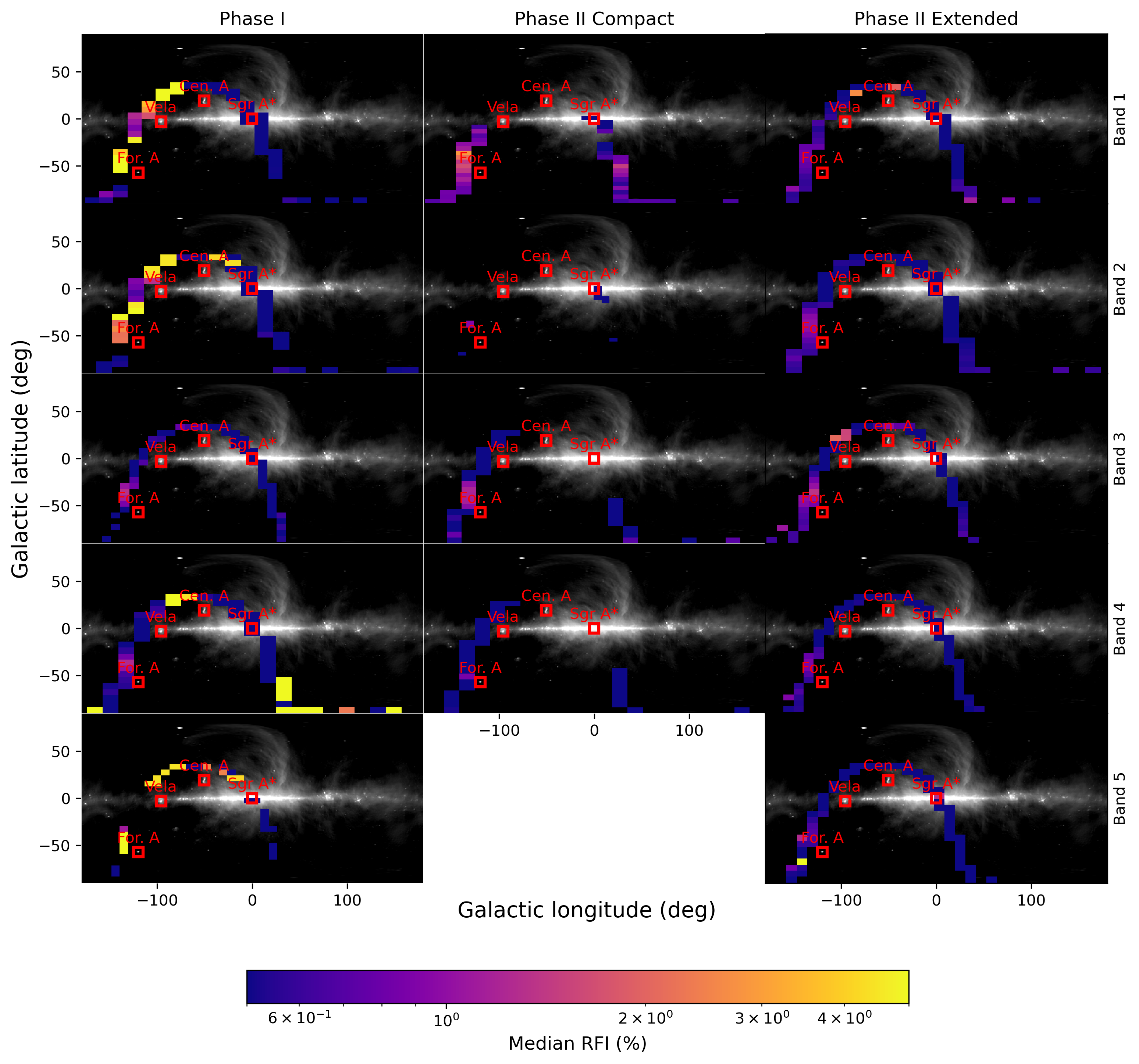}
    \caption{Figure shows the median percentage of data flagged by {\tt AOFLAGGER}, using the default flagging strategy, as a function of Galactic pointing for different frequency-band and array-configuration combinations. The locations of several bright radio sources---Vela, Fornax~A (For.~A), Centaurus~A (Cen.~A), and Sagittarius~A* (Sgr~A*)---are marked by red boxes.}
    \label{fig:mwaBL}
\end{figure*}

% \noindent\textbf{Etc.}

% \iacunnumbered{References}
% \textit{In the text}

% Indicate references by number(s) in square brackets in line with the text. The actual authors can be referred to, but the reference number(s) must always be given. Example: ``\ldots\ as demonstrated [3,6]. Barnaby and Jones [8] obtained a different result \ldots''

% \vspace{5pt}
% \noindent\textit{List of references}

% Number the references (numbers in square brackets) in the list in the order in which they appear in the text. Examples:

\end{document}